\documentclass[conference]{IEEEtran}
\usepackage{amsmath,amssymb,amsfonts}
\usepackage{algorithmic}
\usepackage{graphicx}
\usepackage{biblatex}
\usepackage{textcomp}
\usepackage{xcolor}
\usepackage{subcaption}

\def\BibTeX{{\rm B\kern-.05em{\sc i\kern-.025em b}\kern-.08em T\kern-.1667em\lower.7ex\hbox{E}\kern-.125emX}}
\begin{document}

\title{Comparison of ConvNeXt and Vision-Language Models for Breast Density Assessment in Screening Mammography}

\newcommand{\linebreakand}{%
  \end{@IEEEauthorhalign}
  \hfill\mbox{}\par
  \mbox{}\hfill\begin{@IEEEauthorhalign}
} 

\author{
\IEEEauthorblockN{1\textsuperscript{st} Yusdivia Molina-Roman}
\IEEEauthorblockA{\textit{School of Engineering and Sciences} \\
\textit{Tecnologico de Monterrey}\\
Mexico City, Mexico \\
ORCID: 0009-0005-3623-8971}
\and
\IEEEauthorblockN{2\textsuperscript{nd} David Gómez-Ortiz}
\IEEEauthorblockA{\textit{Centro de Tecnología Biomédica,} \\
\textit{Universidad Politécnica de Madrid} \\
Madrid, Spain \\
da.gomez@upm.es\\
ORCID: 0009-0003-3246-5435}
\and
\IEEEauthorblockN{3\textsuperscript{rd} Ernestina Menasalvas-Ruiz}
\IEEEauthorblockA{\textit{Centro de Tecnología Biomédica} \\
\textit{Universidad Politécnica de Madrid} \\
Madrid, Spain \\
ernestina.menasalvas@upm.es\\
ORCID: 0000-0002-5615-6798}
\and
\IEEEauthorblockN{4\textsuperscript{th} José Gerardo Tamez-Peña}
\IEEEauthorblockA{\textit{School of Medical and Health Sciences} \\
\textit{Tecnologico de Monterrey}\\
Monterrey, Mexico \\
ORCID: 0000-0003-1361-5162}
\and
\IEEEauthorblockN{5\textsuperscript{th} Alejandro Santos-Diaz}
\IEEEauthorblockA{\textit{School of Engineering and Sciences} \\
\textit{Tecnologico de Monterrey}\\
Mexico City, Mexico \\
ORCID: 0000-0001-5235-7325}
}
\maketitle

\begin{abstract}

Mammographic breast density classification is essential for cancer risk assessment but remains challenging due to subjective interpretation and inter-observer variability. This study compares multimodal and CNN-based methods for automated classification using the BI-RADS system, evaluating BioMedCLIP and ConvNeXt across three learning scenarios: zero-shot classification, linear probing with textual descriptions, and fine-tuning with numerical labels.
Results show that zero-shot classification achieved modest performance, while the fine-tuned ConvNeXt model outperformed the BioMedCLIP linear probe. Although linear probing demonstrated potential with pretrained embeddings, it was less effective than full fine-tuning. These findings suggest that despite the promise of multimodal learning, CNN-based models with end-to-end fine-tuning provide stronger performance for specialized medical imaging. The study underscores the need for more detailed textual representations and domain-specific adaptations in future radiology applications.

\end{abstract}

\section{Introduction}
Accurate breast density classification plays a critical role in assessing breast cancer risk. High breast density has been shown to both obscure tumor detection on mammograms and correlate with an elevated risk of developing breast cancer \cite{ McCormack2006}. As a result, the precise evaluation of breast density is essential for early diagnosis and appropriate clinical management.

Manual classification of mammographic density remains a complex and subjective task. Mammograms can be difficult to interpret due to overlapping tissue structures, and assessments often rely heavily on the visual judgment of radiologists. The digitization of medical imaging has opened the door to computational methods capable of reducing variability and improving consistency. Deep learning approaches, particularly convolutional neural networks (CNNs), have emerged as powerful tools. Nevertheless, these models often require large amounts of labeled data and are prone to overfitting, especially in complex domains like mammography. Consequently, a careful balance between automated systems and human expertise is essential for achieving clinically reliable outcomes.



Beyond vision-only models, multimodal learning approaches that combine image and text data have gained attraction in the medical domain. These models leverage the information available in electronic health records (EHRs) and radiology reports to enhance decision-making. Studies have shown that multimodal AI can outperform unimodal counterparts in a range of biomedical tasks by improving data efficiency and contextual understanding \cite{schouten2024}. In the context of breast density assessment, vision-language models (VLMs) offer the opportunity to utilize accompanying clinical text—such as radiologist reports—to improve classification performance and interpretability.

In this work, we address the task of breast density classification according to the BI-RADS (Breast Imaging-Reporting and Data System) density scheme \cite{birads_reference} and leveraging a dataset of annotated mammographic images and corresponding radiology reports collected from the San José Hospital at TecSalud, Tecnológico de Monterrey, in Monterrey, Mexico. We conduct a comparative analysis of two state-of-the-art approaches: ConvNeXt, a CNN-based deep learning model \cite{liu2022}, and BioMedCLIP, a VLM pretrained with token-based textual labels \cite{zhang2025biomedclip}. The main contribution of this work is to compare a VLM and a CNN-based model for the task of breast density classification using paired mammographic images and radiology reports. 



\section{Related work}
\label{sec:RelatedWork}

Breast density has been evaluated through various approaches, including traditional machine learning, image-based deep learning, and more recently, multimodal VLMs. 

Early approaches to breast density classification relied on traditional machine learning methods such as Support Vector Machines (SVMs), using handcrafted statistical and textural features. High accuracies were reported—up to 97\% \cite{Arefan_2015}—and pipelines combining preprocessing with classifiers like random forests further improved performance \cite{mehrabi_2025}. However, these methods depend heavily on expert-designed features, which are time-consuming to create and subject to variability \cite{alhusari_2025}. As a result, their generalization in clinical settings remains a challenge. 

Deep learning has greatly advanced breast density estimation by eliminating the need for manual feature extraction. CNNs have shown strong performance in capturing complex mammographic patterns \cite{inoue_2021}, while transformer-based models have demonstrated potential in related medical imaging tasks \cite{SHAMSHAD_2023} thanks to their ability to model global context \cite{alhusari_2025, jiang_2024}. However, these models still face limitations including high data and computational requirements.

ConvNeXt, a refined version of ResNet-50, combines the efficiency of CNNs with the performance of transformers \cite{liu2022}. Its strong results on ImageNet and its ability to leverage transfer learning make it well-suited for medical imaging tasks with limited labeled data. Studies have confirmed its scalability and accuracy in domain-specific applications, including breast density estimation \cite{Agastya_Todi_2023, YangZujian_2022}.

VLMs have shown strong potential in medical imaging by aligning visual and textual information through large-scale pretraining. CLIP \cite{radford2021CLIP} and its medical adaptations—PubMedCLIP \cite{eslami2023-pubmedclip} and BioCLIP \cite{Stevens2024}—demonstrated improved performance in medical tasks, with domain-specific pretraining yielding notable gains. In breast imaging, Mammo-CLIP \cite{ghosh_2024} achieved robust classification and localization of mammographic features, highlighting the promise of VLMs for enhancing accuracy and generalization in clinical applications.

BioMedCLIP \cite{zhang2025biomedclip} is a VLM tailored for biomedical applications, pretrained on over 15M image-caption pairs from PubMed Central. Using a frozen text encoder and a contrastive-trained image encoder, visual features are aligned with clinical semantics. This enables effective image embeddings for downstream tasks such as classification, retrieval, and visual question answering, showcasing the model’s ability to bridge visual data with domain-specific knowledge.

Collectively, these works illustrate the growing relevance of VLMs in biomedical imaging, particularly in settings where annotated data is limited and semantic alignment between text and image enhances task performance. Despite the promising capabilities of VLMs, it remains unclear whether they consistently outperform conventional convolutional or transformer-based vision models in breast density estimation and related clinical tasks. While VLMs offer advantages in multimodal reasoning and semantic alignment, their effectiveness is highly dependent on the quality and relevance of pretraining data, as well as task-specific fine-tuning. 

\section{Methodology}
\label{sec:Materials}
\subsection{Data preprocessing}
\begin{figure}[htb!]
    \centering
    \includegraphics[width = 6.5 cm]{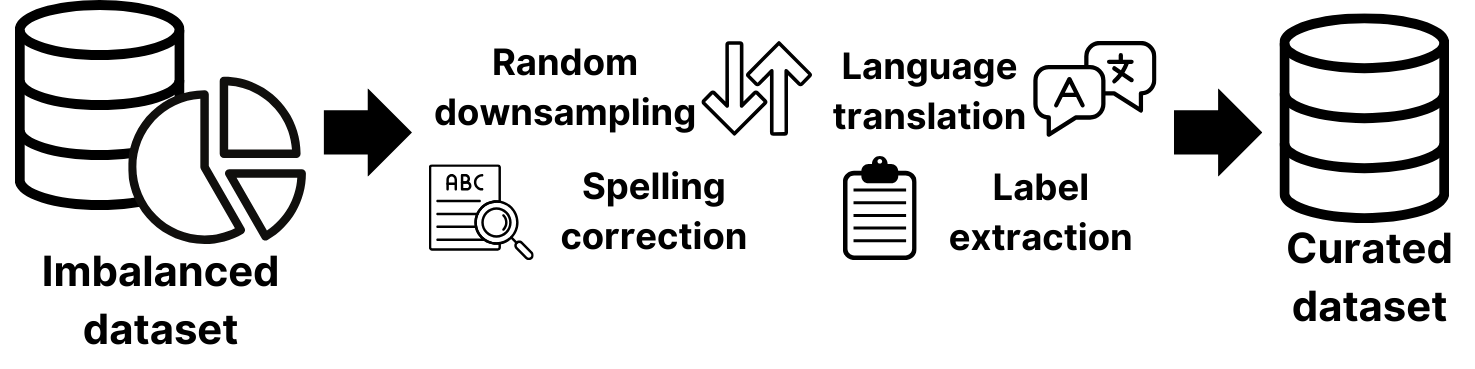}
    \caption{Outline of the processes applied to dataset.} 
    \label{fig:esquema_data}
\end{figure}

This study uses a comprehensive dataset collected from the San Jose Hospital at TecSalud, Tecnológico de Monterrey, in Monterrey, Mexico. The dataset underwent rigorous data cleaning and labeling procedures to ensure its integrity, following strict security and privacy protocols established by TecSalud.\footnote{The institutional ethics board approved the study.}

The dataset comprises electronic health records (EHRs) spanning from 2014 to 2019, encompassing $1,160$ cases. Each case corresponds to a screening mammography exam and includes two standard mammographic views—mediolateral oblique (MLO) and craniocaudal (CC)—for both breasts, resulting in a total of $4,640$ images paired with $1,160$ unique text reports.

An overview of the processes applied to the original dataset can be seen in Fig. \ref{fig:esquema_data}. The original radiology reports, written in Spanish, included clinical indications, imaging findings, and a diagnostic conclusion. These reports often contained textual inconsistencies, such as misspellings, vowel substitutions, and irregular spacing. Following the preprocessing methodology proposed by \cite{Salazar_2024}, these issues were corrected and the reports were subsequently translated into English. Breast density information was subsequently extracted from the findings section using regular expressions. To ensure consistency, the extracted statements were standardized into four BI-RADS-compliant categories \cite{birads_reference}. Reports without a clear density classification were excluded from the final dataset. The resulting class distribution was as follows: 
\begin{itemize}
    \item \textit{Heterogeneously dense}: $1,796$ images
    \item \textit{Scattered areas of fibroglandular density}: $792$ images
    \item \textit{Extremely dense}: $788$ images
    \item \textit{Fatty predominance}: $440$ images
\end{itemize}

Mammographic images were contrast enhanced via a histogram matching process described in \cite{bosques2024} to minimize inter-device variability. The class distribution was initially imbalanced, with \textit{Fatty predominance} representing the least frequent category, totaling only $440$ cases. To mitigate this imbalance, random downsampling was performed across all categories, resulting in a balanced dataset with approximately $450$ images per breast density class. This curated dataset serves as the basis for a comparative analysis of ConvNeXt and BioMedCLIP, allowing the evaluation of their respective performance in breast density classification using images and radiological report data.

\subsection{Trained models}
\label{sec:Methods}
\begin{figure}[htb!]
    \centering
    \includegraphics[width = 6.5 cm]{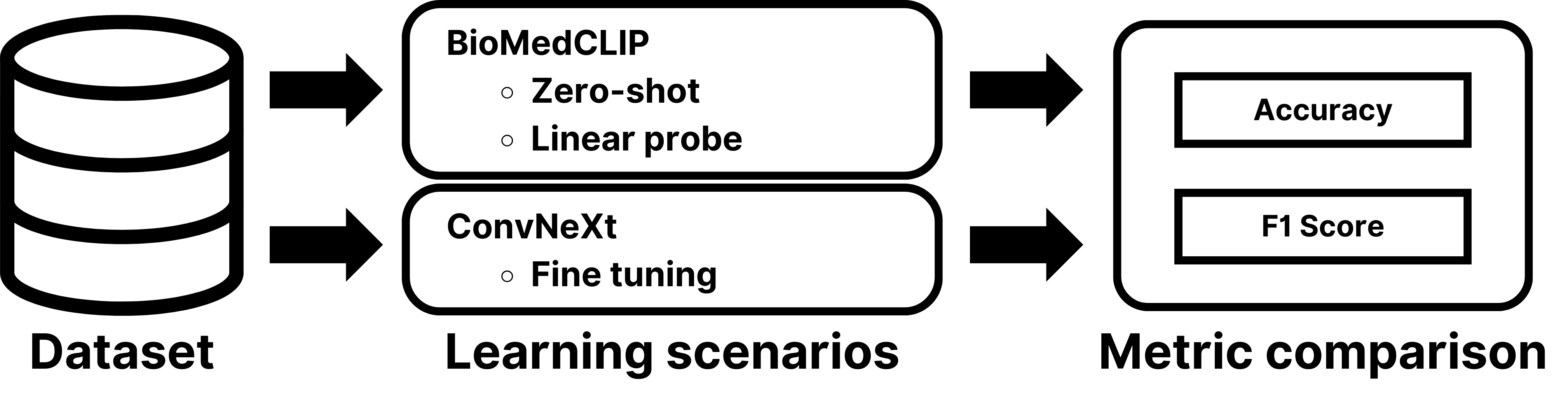}
    \caption{Outline of the methodology used.} 
    \label{fig:esquema_methods}
\end{figure}

This study conducts a comparative analysis of two state-of-the-art models for breast density classification: BioMedCLIP and ConvNeXt. The goal is to evaluate their performance under consistent experimental conditions using a balanced dataset.

\subsubsection{BioMedCLIP: Vision-Language Model}

In this study, BioMedCLIP is evaluated under two learning scenarios:
\begin{itemize}
    \item \textbf{Zero-shot learning:} Classification is performed directly using the pretrained model without any additional training on the target dataset. This setting leverages the model's generalization ability from large-scale pretraining.
    
    \item \textbf{Few-shot learning via linear probing:} The model’s pretrained weights are kept frozen, and a linear classification layer is trained on top of the image embeddings using labeled examples from the breast density dataset. This approach is computationally efficient and requires fewer examples per class compared to full fine-tuning.
\end{itemize}

Linear probing was chosen over fine-tuning for three main reasons: (1) the dataset is relatively small, (2) linear probing is less computationally demanding, and (3) it aligns with the evaluation setup used in the original BioMedCLIP benchmark experiments \cite{patel_2024}.

\subsubsection{ConvNeXt: Vision-Based Model}

ConvNeXt is fine-tuned on the breast density dataset using standard supervised learning. 

\subsubsection{Experimental Setup and Evaluation}

To ensure a fair comparison between the two models, all experiments are conducted using the same dataset. Each of the four density categories is encoded numerically.

Model performance for all the experiments is evaluated using standard classification metrics, including accuracy and F1-score. In addition, confusion matrices are generated to provide a detailed view of classification behavior across the four classes.

\subsection{Experiments}
\label{sec:Experiments}

To evaluate the effectiveness of BioMedCLIP and ConvNeXt we conducted three experiments: zero-shot inference with BioMedCLIP, linear probing with BioMedCLIP, and full fine-tuning with ConvNeXt. Each experiment follows a consistent setup with clearly defined dataset splits, training protocols, and evaluation metrics.

\subsubsection{Experiment 1: BioMedCLIP Zero-Shot Classification}
\begin{itemize}
    \item \textbf{Objective:} This experiment evaluates BioMedCLIP’s performance in a zero-shot setting.
    \item \textbf{Dataset and Evaluation:} Since zero-shot classification does not require model training, the entire dataset is used for inference and evaluation.
    \item \textbf{Model Configuration:} Mammographic images are presented to the pretrained BioMedCLIP model alongside four textual prompts, each corresponding to one of the breast density categories.
    \item \textbf{Training Details:} No training or fine-tuning is performed in this setting.
    \item \textbf{Evaluation Protocol:} The model is evaluated using accuracy, F1-score, and confusion matrix metrics computed over the full dataset.
\end{itemize}
\subsubsection{Experiment 2: BioMedCLIP with Linear Probing}
\begin{itemize}
    \item \textbf{Objective:} This experiment investigates the performance of BioMedCLIP in a few-shot learning scenario using linear probing.
    \item \textbf{Dataset and Splits:} The dataset is split into $85\%$ training and $15\%$ test sets. The training set is further divided into $85\%$ training and $15\%$ validation subsets for hyperparameter tuning and early stopping.
    \item \textbf{Model Configuration:} The pretrained BioMedCLIP encoder is used as a frozen feature extractor. A linear classification head is trained on top of the image embeddings to predict the four breast density categories. The linear layer is initialized using Xavier initialization.
    \item \textbf{Training Details:} Training is conducted using the AdamW optimizer with a learning rate of $0.0001$, a batch size of 64, and a maximum of $200$ epochs. $L_2$ regularization with a weight decay factor of $0.001$ is applied to reduce overfitting and improve generalization.
    \item \textbf{Evaluation Protocol:} Model performance is evaluated on the held-out $15\%$ test set using accuracy, F1-score, and confusion matrices.
\end{itemize}

\subsubsection{Experiment 3: ConvNeXt Fine-Tuning}
\begin{itemize}
    \item \textbf{Objective:} This experiment benchmarks ConvNeXt, a vision-only model, by fine-tuning it end-to-end for breast density classification.
    \item \textbf{Dataset and Splits:} The dataset is split identically to Experiment 2.
    \item \textbf{Model Configuration:} A ConvNeXt-Base model pretrained on ImageNet is used. Its final classification head is replaced with a new dense layer adapted to the four breast density classes. The entire network is fine-tuned during training.
    \item \textbf{Training Details:} Training is performed using the AdamW optimizer with a learning rate of $0.0001$, a batch size of $64$, and a maximum of $200$ epochs. Early stopping is used to stop training if the validation loss shows no improvement for $10$ consecutive epochs, with convergence usually occurring around epoch $40$.
    \item \textbf{Evaluation Protocol:} The final model is evaluated on the same $15\%$ test set as BioMedCLIP, using accuracy, F1-score, and confusion matrices for comparison.
\end{itemize}

\section{Results}
\label{sec:Results}
\begin{table}[htbp]
    \caption{Model performance for breast density classification. \label{tab1:results}}
    \begin{center}
        \begin{tabular}{|c|c|c|}
            \hline
            \textbf{\textit{Model}} & \textbf{\textit{Accuracy}} & \textbf{\textit{F1 Score}}\\
            \hline
            \textit{BioMedCLIP (zero-shot)} & $0.47$ & $0.31$ \\
            \hline
            \textit{BioMedCLIP (linear probe)} & $0.64$ & $0.63$ \\
            \hline
            \textit{ConvNeXt fine-tune} & \textbf{$0.73$} & \textbf{$0.78$} \\
            \hline
        \end{tabular}
    \end{center}
\end{table}

An overview of the results obtained for the three learning scenarios can be seen in Table \ref{tab1:results}.

\subsection{Zero-Shot Classification}
Zero-shot classification approach using BioMedCLIP aims to classify each mammogram into one of four categories without additional task-specific training. This approach obtained an accuracy of $0.47$ and a F1 score of $0.31$.

\subsection{Linear Probing}
Introducing a new layer on top of the frozen BioMedCLIP image encoder significantly improved classification performance, reaching an accuracy of $0.64$ and a F1 score of $0.63$. 

\begin{figure}[htb!]
    \centering
    \includegraphics[width = 6.5 cm]{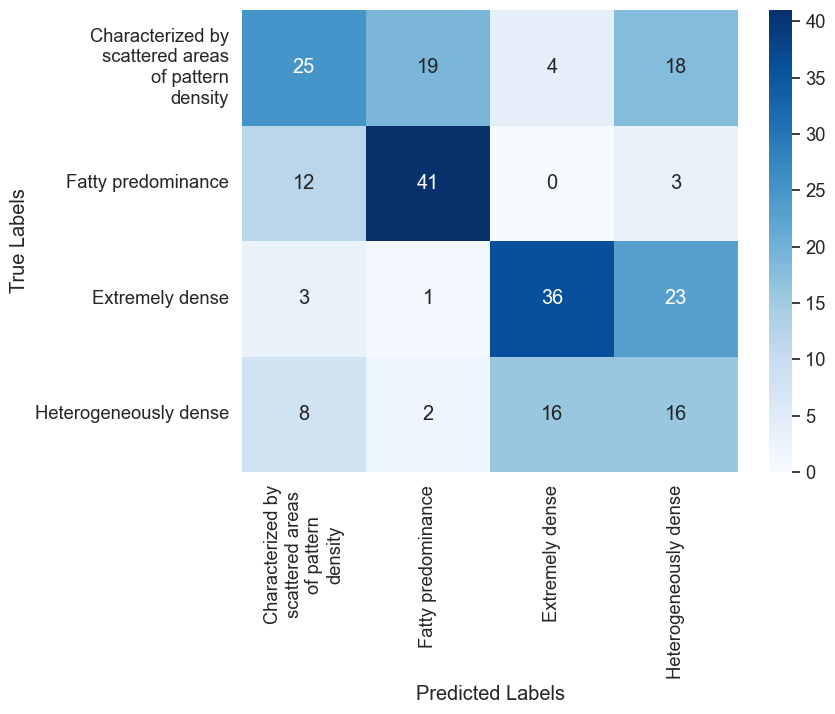}
    \caption{Confusion matrix for BioMedCLIP with linear probing.}
    \label{fig:confusion_biomedclip}
\end{figure}

The per-class validation accuracy ranged from $0.51$ to $0.83$, where the category with the highest performance is \textit{Fatty predominance} and the most challenging category to identify is \textit{Heterogeneously dense}, as shown in the confusion matrix in Figure \ref{fig:confusion_biomedclip}.

\subsection{Fine-Tuning}
\begin{figure}[htb!]
    \centering
    \includegraphics[width = 6.5 cm]{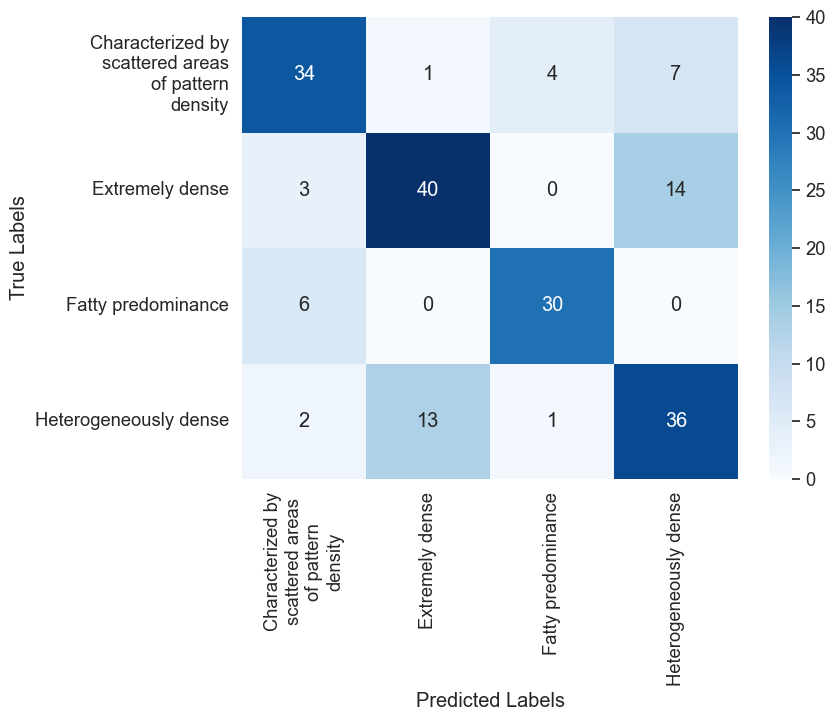}
    \caption{Confusion matrix for ConvNeXt fine-tune.}
    \label{fig:confusion_convnext}
\end{figure}

Fine-tuning the ConvNeXt base model yields the best results among the three learning scenarios. It achieves a validation accuracy of $0.73$. The validation accuracy per class ranges between $0.58$ and $0.82$, with the highest accurately predicted category being \textit{Extremely dense} and the most challenging category being \textit{Characterized by scattered areas of pattern density}. The validation F1 score values per class range from $0.6$ to $0.78$; the highest values obtained for the \textit{Extremely dense} and  \textit{fatty predominance} categories, and \textit{Heterogeneously dense} being the most difficult class to identify.

\section{Discussion} 
\label{sec:Discussion}

\textbf{Zero-Shot Performance of BioMedCLIP.} The zero-shot application of BioMedCLIP achieved an accuracy of $0.47$ but suffered from a low average F1-score of $0.31$, revealing a significant class imbalance in its predictions. Despite the advantages of large-scale multimodal pretraining, the model struggled to interpret the specific visual features and terminologies associated with mammographic density. Without domain-specific tuning, BioMedCLIP had difficulty linking mammographic patterns to the corresponding textual descriptions, highlighting a key limitation of using VLMs in specialized medical imaging tasks.

This underperformance reinforces the broader challenge of transferring general biomedical representations to specialized diagnostic fields like breast imaging. Consistent with prior research, these results emphasize the need for domain-specific adaptations to optimize performance in medical applications. While zero-shot evaluation can provide a baseline for assessing robustness and generalization, it remains inadequate for critical clinical tasks such as breast density classification.

\textbf{Linear Probing Performance of BioMedCLIP.} Training a linear classifier on BioMedCLIP's pretrained image-text embeddings significantly improved classification performance compared to the zero-shot approach. While the model performed well, it struggled with the \textit{Heterogeneously dense} class, achieving an F1 score of $0.52$. This suggests that while the model’s latent features contain useful information, they may lack the fine-grained specificity needed to differentiate this more ambiguous category reliably. 

Analysis of the confusion matrix reveals that the model effectively distinguished between the density extremes—\textit{Fatty predominance} and \textit{Extremely dense}—with its highest accuracy recorded in the former. However, it had difficulty classifying intermediate categories like \textit{Scattered} and \textit{Heterogeneously dense}, which exhibit lower recall due to subtle textural differences. This pattern of confusion reflects known challenges in breast density classification, even for human experts. These findings highlight that while BioMedCLIP's pretrained embeddings capture relevant semantic features, incorporating a task-specific classification layer through linear probing is crucial for adapting them to the complexities of mammographic image interpretation.

\textbf{Fine-Tuning Performance of ConvNeXt.} The ConvNeXt model, when fine-tuned end-to-end on the breast density dataset, outperformed all other evaluated approaches in terms of accuracy and F1-score. By fully leveraging its feature extraction capacity, ConvNeXt could learn a direct numeric mapping of breast density classes, leading to a more consistent and balanced classification performance than BioMedCLIP’s linear probing strategy. The model particularly excelled in distinguishing the \textit{Fatty predominance} and \textit{Extremely dense} categories, where visual features are more pronounced, though it faced challenges with the more ambiguous \textit{Scattered} and \textit{Heterogeneously dense} categories.

An analysis of the confusion matrix showed in ~\ref{fig:confusion_convnext} highlighted that opposing categories, such as \textit{Fatty predominance} and \textit{Extremely dense}, were rarely misclassified due to their distinct visual features. However, significant confusion remained between adjacent categories, especially with \textit{Scattered}, which was frequently mistaken for both \textit{Fatty predominance} and \textit{Heterogeneously dense}. ConvNeXt performed best in identifying \textit{Fatty predominance}, whereas \textit{Heterogeneously dense} tissues remained the most difficult to classify due to their subtle and overlapping visual characteristics.

While ConvNeXt demonstrated strong performance through end-to-end fine-tuning, it still struggled with breast density categories that lie close together on the BI-RADS continuum. Comparisons with BioMedCLIP revealed that both models found distinguishing higher-density classes challenging, but ConvNeXt achieved a higher recall for lower-density categories, particularly \textit{Scattered} areas. These findings emphasize the advantages of domain-specific fine-tuning in improving classification reliability and suggest that further architectural enhancements or training strategies may be needed to address remaining classification ambiguities.

\textbf{Token-Based vs. Numerical Classification: Challenges and Limitations.} Multimodal representation learning has shown promise in medical imaging but faces challenges due to data heterogeneity and the complexity of medical terminology \cite{schouten2024}. While models like CLIP excel in general computer vision tasks through large-scale image-text pretraining, their effectiveness in specialized medical domains is limited. Zero-shot classification struggles with generic prompts that fail to capture nuanced medical descriptions. Aditionally, CLIP’s dual-encoder architecture can introduce representational gaps between visual and textual modalities, reducing diagnostic accuracy \cite{liu_2025}.

A major barrier to applying VLMs in medical imaging is the lack of large, high-quality annotated datasets for contrastive pretraining. Without sufficient domain-specific data, these models fail to generalize well across different imaging modalities. To address these limitations, researchers emphasize the need for domain-adapted architectures, carefully curated datasets, and improved prompt engineering strategies. Enhancing alignment between medical images and textual descriptions is crucial for improving model performance in clinical applications.

One potential solution is the use of descriptive tokens or contextual prompts to refine model attention. Studies suggest that aligning text tokens with specific image regions enhances pathology detection, while token labeling in vision transformers improves classification accuracy. However, balancing token granularity is essential, as overly complex token assignments can increase computational costs without significant diagnostic benefits. In experiments, BioMedCLIP’s linear probe struggled with mammographic density classification due to insufficiently detailed textual tokens, as minor wording differences failed to create clear semantic distinctions. These findings highlight the importance of carefully engineered prompts and enriched token representations when adapting VLMs to specialized medical tasks.

\section{Conclusions and future work}
\label{sec:Conclusions}
This study compared CNN-based architectures with VLMs for breast density classification, revealing that while multimodal approaches hold promise, they face challenges in specialized medical tasks without proper adaptation. The results showed that fine-tuned ConvNeXt consistently outperformed BioMedCLIP in accuracy and F1-score, with the latter struggling particularly in zero-shot settings. These findings highlight the necessity of domain-specific fine-tuning, even for advanced pre-trained models, to ensure reliable clinical performance.

One key limitation of BioMedCLIP stemmed from the lack of granularity in textual descriptors defining breast density categories. The model struggled to distinguish subtle linguistic variations--such as "extremely" versus "heterogeneously"--leading to weaker visual-text alignments. This emphasizes the need for carefully designed prompts and more descriptive textual tokens to enhance multimodal learning. Future research should focus on improving textual representations and exploring domain-specific pretraining or adaptive fine-tuning to address these challenges.

Ultimately, this study contributes to AI-driven radiology by highlighting both the limitations and potential of VLMs. While CNN-based architectures currently achieve superior performance, multimodal approaches offer valuable interpretability and flexibility. Advancing these models with enriched textual representations and adaptive learning strategies could help bridge the performance gap, paving the way for more semantically grounded and clinically useful medical imaging systems.

\section{Acknowledgments}

The authors would like to thank TecSalud for providing the clinical data used in this study. This work was supported by Tecnológico de Monterrey and CONAHCYT (grant number 1317813), as well as by ELADAIS (https://eladais.org/), funded by the Spanish Ministry of Economic Affairs and Digital Transformation under the UNICO I+D Cloud program.

\printbibliography

\end{document}